# Gate-tunable van der Waals heterostructure for reconfigurable neural network vision sensor

Chen-Yu Wang[1,†], Shi-Jun Liang[1,†], Shuang Wang[1], Pengfei Wang[1], Zhu'an Li[1], Zhongrui Wang[2], Anyuan Gao[1], Chen Pan[1], Chuan Liu[3], Jian Liu[3], Huafeng Yang[3], Xiaowei Liu[1], Wenhao Song[2], Cong Wang[1], Xiaomu Wang[3], Kunji Chen[3], Zhenlin Wang[1], Kenji Watanabe[4], Takashi Taniguchi[4], J. Joshua Yang[2,*] & Feng Miao[1,*].

[1]National Laboratory of Solid State Microstructures, School of Physics, Collaborative Innovation Center of Advanced Microstructures, Nanjing University, Nanjing, 210093, China.

[2]Department of Electrical and Computer Engineering, University of Massachusetts, Amherst, MA, USA.

[3]School of Electronic Science and Engineering, Nanjing University, Nanjing, 210093, China.

[4]National Institute for Materials Science, 1-1 Namiki Tsukuba, Ibaraki, 305-0044, Japan.

[†]These authors contributed equally to this work: Chen-Yu Wang and Shi-Jun Liang

[*]Corresponding author: Email: jjyang@umass.edu; miao@nju.edu.cn.

**Early processing of visual information takes place in the human retina. Mimicking neurobiological structures and functionalities of the retina provide a promising pathway to achieving vision sensor with highly efficient image processing. Here, we demonstrate a prototype vision sensor that operates via the gate-tunable positive and negative photoresponses of the van der Waals (vdW) vertical heterostructures. The sensor emulates not only the neurobiological functionalities of bipolar cells and photoreceptors but also the unique synaptic connectivity between bipolar cells and photoreceptors. By tuning gate voltage for each pixel, we achieve reconfigurable vision sensor for simultaneously image sensing and processing. Furthermore, our prototype vision sensor itself can be trained to classify the input images, via updating the gate voltages applied individually to each pixel in the sensor. Our work indicates that vdW vertical heterostructures offer a promising platform for the development of neural network vision sensor.**

Traditional vision chips (*1*) separate image sensing and processing, which would limit their performance with increasing demand for real-time processing (*2*). In contrast, the human retina has a hierarchical biostructure for connectivity among neurons with distinct functionalities and enables simultaneous sensing and preprocessing of visual information. A principal function of the human retina is to extract key features of the input visual information by preprocessing operations, although the specific neuronal activities remain a subject of intensive investigations (*3-5*). This function aims to discard the redundant visual data and substantially accelerates further information processing in the human brain, such as pattern recognition and interpretation (*6*). Therefore, implementing retinomorphic vision chips represent a promising solution to solve the challenge faced by traditional chips and to process a large volume of visual data in practical applications (*7, 8*). So far, various technologies have been proposed to emulate the functions of the retina to integrate the image sensor and processing unit in each pixel for retinomorphic applications (*2, 8-12*). Alternative to these conventional technologies, optoelectronic resistive random access memory synaptic devices allow for achieving the functions of image sensing and preprocessing as well as memory (*13*), showing promise in reducing the complex circuitry for artificial visual system. To meet the increasing demands for edge computing, developing more advanced image sensors, such as with reconfigurable and self-learning capabilities, is highly desirable. Exploiting novel physical phenomena of emerging atomic-scale materials and hierarchical architectures made of these materials may offer a promising approach to realize such neural network vision sensors.

Two-dimensional (2D) materials with atomic thickness and flatness have shown great potential for numerous applications in electronics (*14-17*) and optoelectronics (*18-20*). The vdW vertical heterostructures formed by stacking different 2D materials accommodate an abundance of electronic and optoelectronic properties (*21-32*), which may be exploited to mimic hierarchical architecture and functions of retinal neurons (*33-43*) in a natural manner to implement a neural network vision sensor. Here, we show that the image sensor based on vdW vertical heterostructures can emulate the biological characteristics of retinal microcircuits made of photoreceptors, bipolar cells and the synaptic connectivity between photoreceptors and bipolar cells. Besides, the fabricated vision sensor can be programmed to simultaneously sense images and process them with distinct kernels. Most importantly, we also demonstrated that the retinomorphic vision sensor itself is capable of being trained to carry out the task of pattern recognition. The technology proposed in this work opens up opportunities for the implementation of advanced vision chips in the future.

As mentioned above, different types of retinal neurons are organized in a hierarchical way (Fig.

1a). More than 50 types of cells are distributed within a few different layers in the vertebrate retina, such as the photoreceptor layer, bipolar cell layer and ganglion cell layer (*44*). The layered structure creates various types of retinal microcircuits that constitutes the distinct visual pathways in the retina and ensures the information flows from the top to the bottom. In these microcircuits, cone cells (one type of photoreceptor) and bipolar cells are crucial neurons. The cone cells transduce visual signals into electrical potential, while the bipolar cells serve as the critical harbors for shaping input signals (Fig. 1a), which can accelerate perception in the brain. According to their distinct response polarities, bipolar cells can be classified into ON cells and OFF cells, which respond to a light stimulus in opposite manners. Under a light stimulus, glutamate release from the cones is suppressed. The ionic channels of OFF cells with ionotropic glutamate receptors are closed due to the lack of glutamate to attach to. The resulting hyperpolarized OFF cells reduce the membrane potential (green curve in Fig. 1b). Conversely, suppressing the glutamate release by light stimulus opens the ionic channels of ON cells with metabotropic glutamate receptors. The resulting depolarized ON cells show an increased membrane potential (red curve in Fig. 1b). Through these ON and OFF bipolar cells in the pathway (*45*), information can be preprocessed and relayed to the visual cortex in the brain to be further processed for perception.

**Mimicking retinal cells with vertical heterostructure devices**

To emulate the hierarchical architecture and biological functionalities of photoreceptors and bipolar cells layers, we fabricated $WSe_2$/h-BN/$Al_2O_3$ heterostructure device (Fig. 1c). In contrast to the complex structure of silicon retina, the vdW device architecture in which we vertically integrate photoreceptor and bipolar cells is simple and compact. The device fabrication and Raman characterization of $WSe_2$ can be found in the Methods and Supplementary Fig. 1a. Compared to the emulation of ON photoresponse feature, it is more challenging in mimicking OFF photoresponse characteristic. Prior works have shown that the electrical current of devices can be suppressed by light-induced reduction in carrier mobility of low-dimensional material. However, the resulting response time is incomparable to that of ON photoresponse (*22, 23, 46, 47*). By using atomically-sharp interface of vdW vertical heterostructure and $Al_2O_3$ with nanoscale thickness, we are able to overcome the challenge and achieve fast photoresponse speed. The light-induced change of electrical current ($\Delta I_{ds}$), which represents the photoresponse of the devices, is measured from the source/drain electrodes deposited on the $WSe_2$ channel. The vertical heterostructure devices enable the conversion between light and electric signal and exhibit positive photoresponse (positive $\Delta I_{ds}$) and negative photoresponses (negative $\Delta I_{ds}$) dependent on the gate voltage, resembling the biological characteristics of photoreceptors and bipolar cells. Without applying gate voltage, light illumination generates excess electrons and holes in $WSe_2$ channel to increase the current change (Fig. 1d). This source/drain current increasing ("ON-photoresponse") feature under the light stimulus is similar to light-stimulated increase in the membrane potential of ON bipolar cells. With applying a negative gate voltage, the ambipolar $WSe_2$ is electrostatically doped with holes (*14*) (Supplementary Fig. 1b and Supplementary Fig. 1c), the source/drain current decreasing characteristic ("OFF-photoresponse") resembles the light-stimulated reduction in the membrane potential of OFF bipolar cells. Such OFF photoresponse feature is highly reproducible in devices with similar parameters (Supplementary Fig. 2).

The physical origin of OFF photoresponse can be understood in the following way. The existence of nitrogen vacancies in h-BN has been pointed out by many prior studies of cathodoluminescence and elemental analysis (*22, 48-50*), and consequently confirmed by the scanning tunneling microscopy experiment (*51*). With the light illumination on vdW heterostructure devices, electrons of the donor-like nitrogen vacancies distributed in different layers of h-BN are excited and then migrate upward under the action of perpendicular electric field from the back-gate voltage (*51*). The positively charged nitrogen vacancies distributed in the upper layers of h-BN are recombined by other photogenerated electrons migrated from lower part of h-BN. The positively charged vacancies localized in close proximity to the interface of h-BN and $Al_2O_3$ are not recombined during light illumination, effectively screening the black-gate electric field and suppressing the conduction of $WSe_2$ channel (Supplementary Fig. 3a). This screening effect can be enhanced by increasing the light intensity (Supplementary Fig. 3b) and the photoresponse of the devices is able to operate in the entire visible spectrum (Supplementary Fig. 3c). When the light is removed, the electrons tunneling through the thin $Al_2O_3$ layer (~6 nm thick) would recombine with those positively charged defects, and the reduced current rapidly recovers. We have carried out control experiments (Supplementary Fig. 4a-4c) and carefully ruled out the possibility that trap centers on the surface of oxidation layer cause the OFF photoresponse. Based on the reduction of current upon light illumination, we can estimate the concentration of defects in h-BN to be around $10^{10}$ cm$^{-2}$ (Supplementary Fig. 5). This concentration is comparable to that reported in prior works (*22, 51*), further indicating that the OFF photoresponse arises from electron excitation of the vacancy defects in the h-BN.

The vdW heterostructure devices show good performance in terms of operating speed and power consumption. The sharp vdW interface enables us to achieve response time of less than 8 ms (Supplementary Fig. 6) in OFF-photoresponse device. This time scale is comparable to that of biological bipolar cells (*52*) and is expected to be further improved through optimizing the thickness of h-BN and $Al_2O_3$. Besides the fast response, the OFF-photoresponse device holds promise in the operation of low power consumption (Fig. 1e). Figure 1e presents the current of OFF-photoresponse device at different biases and light intensities. The device is capable of exhibiting OFF-photoresponse at low bias (*e.g.* 10 mV), indicating that the low power consumption is reachable with the device. Compared to the OFF-photoresponse device, ON-photoresponse device exhibits a smaller dark current, resulting in a lower power consumption. Using vertical heterostructure could drastically reduce the complexity in each pixel of conventional retinomorphic circuits. With further optimization on power consumption and operating speed, such vertical devices are promising in emulating more advanced functionalities of human retina.

**Reconfigurable retinomorphic vision sensor**

Assembling these ON and OFF-photoresponse devices into an array (an OFF-photoresponse device in the center surrounded by ON-photoresponse devices) enables the emulation of the biological receptive field (RF). The RF is one of retinal microcircuits indispensable for early visual signal processing. The retinal microcircuits have a center area (green in left panel of Fig. 2a) and surrounding areas (pink in left panel of Fig. 2a). Under a light stimulus, the center and surrounding areas of the biological RF show an antagonistic response, which is characterized by difference-of-Gaussians model (DGM) (Methods). The key role of the RF of bipolar cells in the human retina is to early process visual information by extracting its key features (*53*) to accelerate the visual perception in the brain. We emulate the RF of bipolar cells by integrating 13 vdW heterostructure devices into an array (center panel of Fig. 2a and Supplementary Fig. 7), with individual device controlled by gate voltage separately. According to Kirchhoff's law, output ($I_{output}$) of artificial RF is a summation of photocurrent from all devices ($\sum_{i=0}^{5} \Delta I_{ds}^{i}$), and real-time variations of the output represent the dynamic response to light patterns-changing. With this working principle, the artificial RF can be used for detecting edge of objects, which is a fundamental function of the biological RF. In the experiment, the light was switched on column-by-column to represent a contrast-reversing edge moving from the left to the right side (right panel of Fig. 2a). When the edge moves towards the right side, the current variation increases as more ON-photoresponse devices are activated and peak before the edge reaches the center device. With continuous movement, the OFF-photoresponse device in the center antagonizes the photoresponse of the ON-photoresponse devices, leading to an opposite peak in the photocurrent variation. This behavior can be well described by the widely used DGM model (dashed line in the right panel of Fig. 2a), which also accounts for the dynamic response to the edge moving along other directions (Supplementary Fig. 8). Furthermore, the dynamics response of the device array resembles the behaviors exhibited by the RF of a cat's retinal ganglion cells (Supplementary Fig. 9)

The retinomorphic vision sensor shows the functionality of simultaneous sensing and processing (Fig. 2b), which allows for the implementation of near-data processing. This architecture is completely different from traditional architecture vision chips. As aforementioned, the separation of image sensing and processing in traditional vision chips would reduce the efficiency for processing large amounts of real-time image information (*54*), as all the redundant visual data sensed by cameras have to be first converted to digital data and then transmitted to processors. In contrast, the visual information can be sensed and processed simultaneously by using our retinomorphic vision sensor based on vdW heterostructure without requiring analog-to-digital conversions. As a demonstration, we mapped the difference-of-Gaussian (DoG) kernel (3×3) into our vision sensor by assigning specific values to each gate and realized the edge enhancement of letter "N" (8×8 binary, left panel of Fig. 2c). The experimental details can be found in Methods and Supplementary Fig. 10. The variance of $I_{output}$ in the array is simultaneously recorded (Supplementary Fig. 10b). Reconstructing the data of the current variance yields the experimental and simulated letter "N" (center and right panels of Fig. 2c). The experimental results are in good agreement with the simulation results.

By modulating $V_g$ individually applied to each vdW heterostructure device, we are able to achieve reconfigurable retinomorphic vision sensor to simultaneously sensing images and processing images in three different ways, as shown in Fig. 3. Image stylization refers to the rendering effect that generates a photorealistic or non-photorealistic image. It is mainly implemented by software in computer graphics (*55*). By using the retinomorphic vision sensor, we are able to invert a grayscale image of Nanjing University Logo (Fig. 3b). The stylized image of the Logo is similar to the simulation results. In addition to the function of image stylization, we also use the vision sensor to demonstrate other important functions widely used for image processing, such as edge enhancement and contrast correction, which well reproduce the image features shown in the simulation results. In Fig. 3c, we realize edge enhancement by eliminating the contrast difference between logo patterns (black) and background (white). Furthermore, the contrast in the logo can be corrected by using the vision sensor to display hidden information of the edge due to under/over exposure (Fig. 3d). The detection accuracy of the sensor is not deteriorated by the irregular edge patterns in the logo, which is justified by the good agreement between the experimental and simulation results. These findings indicate that the field of hardware accelerating in image processing may benefit from the use of reconfigurable vision sensor.

**Implementation of a convolutional neural network**

The retinomorphic vision sensor is also promising to form a convolutional neural network and carry out classification task of target images (Fig. 4), in which the weights can be updated by tuning gate voltages applied to each pixel of the vision sensor. We take dot product of the sensed image information and the weights represented by the back gate voltage of each pixel to calculate the total output current. By adopting backpropagation approach, we are able to tune the back gate voltage of each pixel after each epoch. In the experiments, the dataset for training is made of 9 binary figures (3 × 3), including three different types of letters, *i. e.* 'n', 'j' and 'u'. As shown in Fig. 4a, the instruction information representing these figures of letters were input into the retinomorphic vision sensor through laser. Fig. 4b illustrates the training process of the vision sensor for pattern recognition. Initially, all the back gate voltages are set to 0. The modulation of gate voltage ($V_{g_{i,j}}^k$) in each pixel (with $i$ and $j$ representing the pixel location) depends on the feedback of the measured photocurrent for the input $k^{th}$ figure. $f_1$, $f_2$ and $f_3$ correspond to the output of the neural network for three different letters respectively. $\delta^k$ denotes the backpropagated error in the $k^{th}$ iteration. We examine the accuracy of image recognition over training epoch to evaluate the convergence of neural network outputs. As shown in Fig. 4c, the accuracy reaches 100 % with less than 10 epochs, which is obtained by the weighted average of three different convolution kernels (see Supplementary Fig. 11 for details). The inset shows the weight distribution of convolutional neural network vision sensor, corresponding to initial (yellow histogram) and after 10 epochs (blue histogram). To further examine the evolution of the recognized results averaged over all three different types of each specific letter, we present the variation of the output ($f_1$, $f_2$ and $f_3$) of each convolution kernel over the number of training (Fig. 4d). We found that the target letter can be well separated from the input images after 2 epochs. The excellent performance of the prototype vision sensor as a convolutional neural network suggests that the integration of vdW vertical heterostructure may open up a new avenue for achieving highly efficient convolutional neural network for visual processing in a fully analog regime (*56*).

In conclusion, we demonstrate a prototype vision sensor based on vdW heterostructure. This sensor can not only closely mimic the biological functionalities of retinal cells, but also exhibits reconfigurable functions of image processing beyond the human retina. Furthermore, we show that the retinomorphic vision sensor itself can be a convolutional neural network for image recognition. Our work represents a first step towards the development of future reconfigurable convolutional neural network vision sensor.

**Methods**

**Device fabrication**

$WSe_2$ (from HQ Graphene) and h-BN were obtained by a mechanical exfoliation approach. Bottom-gate electrodes (Ti (5 nm)/Au (25 nm)) and source/drain electrodes (Pd (5 nm)/Au (75 nm)) were patterned by standard electrical beam lithography and electrical beam evaporation. $Al_2O_3$ was grown on the bottom-gate electrodes by atomic layer deposition (ALD). We fabricated vdW vertical heterostructures by transferring h-BN and $WSe_2$ onto the substrate with the standard polyvinyl alcohol (PVA) method. We used an atomic force microscope (AFM) to confirm the thickness of $WSe_2$ (2~20 nm), h-BN (10~40 nm) and $Al_2O_3$ (6~10 nm), and robustness of devices is a positive correlation with thickness of materials. Each channel between source/drain electrodes is 1~2 μm and length of $WSe_2$ used is 10~30 μm**.** Before carrying out electrical measurements, all devices were annealed at 573 K in an argon atmosphere for 2 hours to remove photoresist residue.

**Difference-of-Gaussians model**

Difference-of-Gaussians model (DGM) has been widely used to describe the response of the biological RF under a light stimulus, where the output Gaussian function $G_{output}(x,y)$ can be expressed by $G_{output}(x,y) = G_{ON}(x,y) - G_{OFF}(x,y)$. Here, $G_{i=ON/OFF}(x,y) = \frac{1}{2\pi} exp\left(-\frac{1}{2}\left(\left(\frac{x-\mu_{ix}}{\sigma_{ix}}\right)^2 + \left(\frac{y-\mu_{iy}}{\sigma_{iy}}\right)^2\right)\right)$ is a two-dimensional Gaussian function representing the spatial intensity distribution of the responsivity in the center or surround area, where $x$ and $y$ are the space coordinates, $\mu_{ix}$ and $\mu_{iy}$ denote the central coordinate of the biological RF in different areas, and $\sigma_{ix}$ and $\sigma_{iy}$ represent the standard deviation of the spatially distributed photoresponse of the biological RF.

**Electrical measurement**

Each vdW heterostructure device was placed on special PCB in a nitrogen atmosphere. All the devices were connected in parallel via a switch matrix box. We conducted current measurements through a data acquisition card (National Instruments, PCIe-6351) and current amplifier (Stanford Research Systems, Model SR570). The gate voltage was applied by a source measurement unit (Keithley, 2635A). As shown in Supplementary Fig. 9a, two separate measurement channels were involved in the experiment: one for ON-photoresponse devices and another for OFF-photoresponse devices.

**Signal processing**

We acquired current signals from the ON and OFF channels separately. Then, we reduced the noise resulting from $H_2O$ and $O_2$ molecules absorbed on the surface of $WSe_2$ (*57*). This kind of noise would lead to fluctuation of the background current and the current change of ON- and OFF-photoresponse devices over time. To eliminate the effect of noise, we first obtained the fluctuation of the background current under background light illumination; then, we calibrated the photocurrent of ON- and OFF-photoresponse devices by multiplying the weights obtained in the section on pattern generation.

Finally, we added measured current from ON and OFF channels and then reorganized these current data into images. All codes were implemented in the Igor software (Wavemetrics).

**Acknowledgements**

This work was supported in part by the National Natural Science Foundation of China (61625402, 61921005, 61974176), and the Collaborative Innovation Center of Advanced Microstructures and Natural Science Foundation of Jiangsu Province (BK20180330), Fundamental Research Funds for the Central Universities (020414380122, 020414380084). K.W. and T.T. acknowledge support from the Elemental Strategy Initiative conducted by the MEXT, Japan, A3 Foresight by JSPS and the CREST (JPMJCR15F3), JST. Z.W., W.S. and J.Y. were supported in part by TetraMem Inc.

**Author Contributions**

F.M., C.Y.W., S.J.L. and J.J.Y. conceived the idea and designed the experiments. C.Y.W., S. Wang and Z. Li fabricated the device arrays and performed electric experiments on the arrays. C.Y.W. and S.J.L. analyzed experimental data. A.G., C.L. and X.W. assisted in optoelectronic measurement. C.P. assisted in the electrical measurement. J.L., H.Y and K.C. grew $Al_2O_3$ with atomic layer deposition. X.L. and Z.W. performed the measurement of Raman spectrum of $WSe_2$ and h-BN. K.W. and T.T. prepared the h-BN samples. S. W, Z. L. and P. W. implemented the neural network. Z.W., W.S. and C.W. contributed to the discussion for neural network. S.J.L., C.Y.W., F.M. and J.J.Y. wrote the manuscript with inputs from all co-authors.

**Competing financial interests**

The authors declare no competing financial interests.

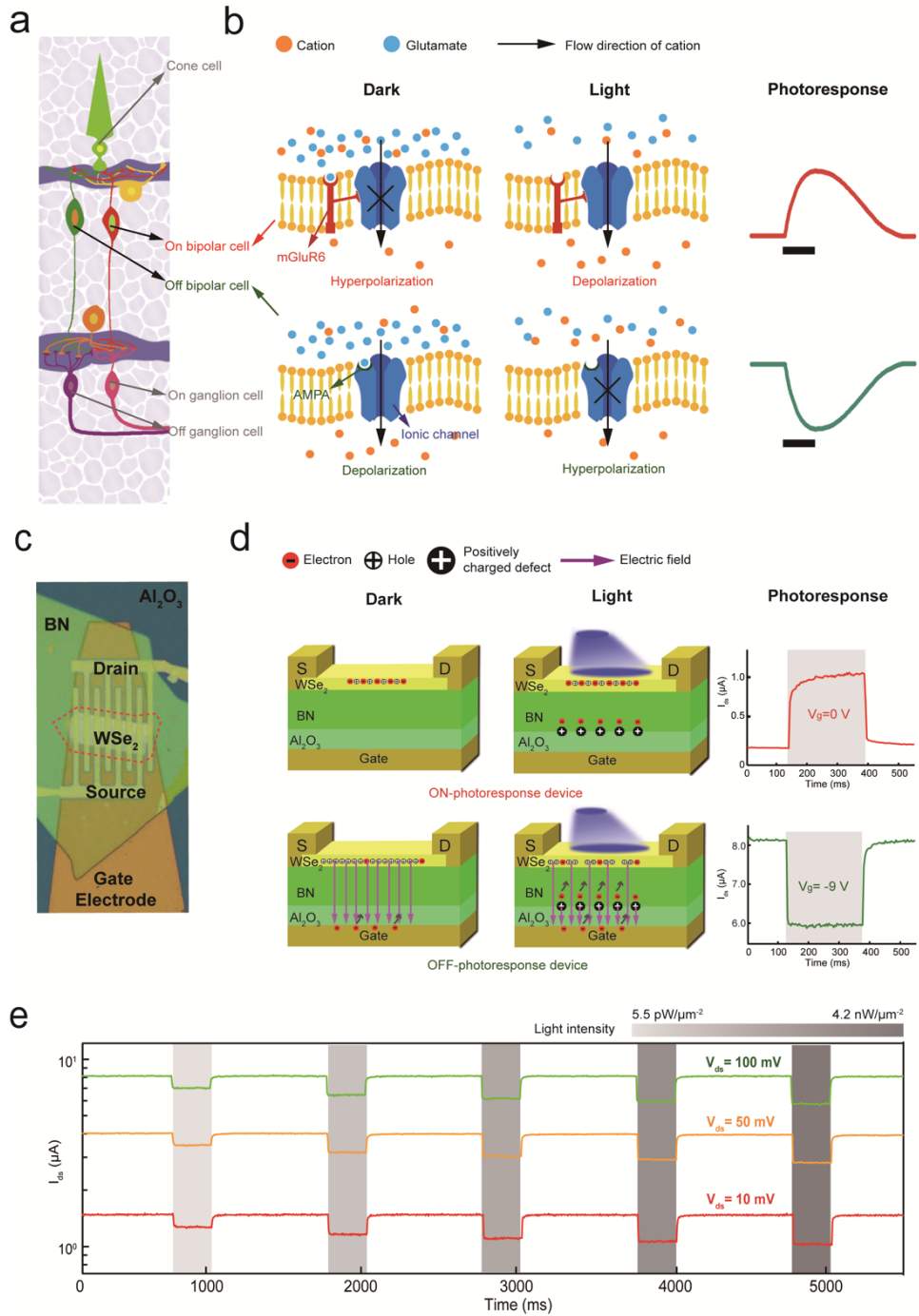

**Fig 1. Retinal and artificial retinal structures**. **a**. The profile of a biological retina. **b**. The biological working mechanism and photoresponse of OFF bipolar cells (with α-amino-3-hydroxy-5-methyl-4-isoxazolepropionic acid, AMPA) and ON bipolar cells (with metabotropic glutamate receptor 6, mGluR6). Black bars in the photoresponse of bipolar cells represent the moment of light illumination. **c**. Optical image of the vdW heterostructure based device. **d.** The operating mechanism and photoresponse of the ON- and OFF-photoresponse devices at zero and negative gate voltages, respectively. The positive (negative) $\Delta I_{ds}$ corresponds to ON- (OFF-) photoresponse. Shadow areas correspond to the duration of light illumination. **e.** OFF-photoresponse at different bias voltages and light intensities (indicated by shadow areas). OFF-photoresponse of the device remains retained at extremely low bias voltage (10 mV), which allows for the operation of low power consumption.

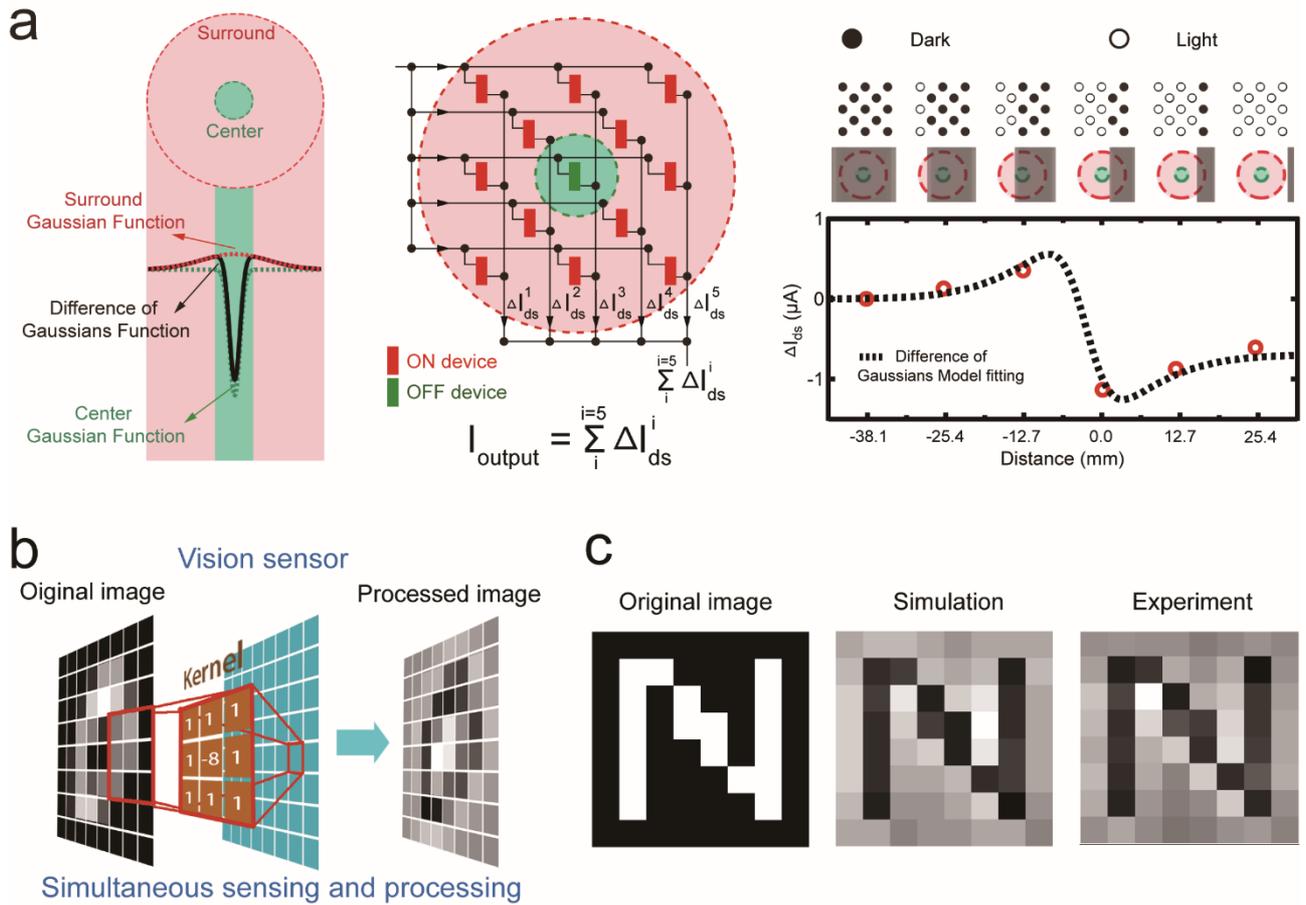

**Fig 2. The retinomorphic vision sensor based on tfvhe vdW vertical heterostructure for simultaneous image sensing and processing.** a. Receptive field (RF) with green center and pink surround areas. Left panel: Difference-of-Gaussians model (DGM) of the RF characterizes the distribution of responsivity; Center panel: vision sensor and its output. An OFF-photoresponse device in the center is surrounded by ON-photoresponse devices. The output of vison sensor is the current summation over all devices; Right panel: outputs of the artificial RF with a contrast-reversing edge moving from the left to the right side. The upper circle array represents light sources. Light is on for solid circles and off for circles. **b**. Vision sensor for simultaneous image processing and processing. **c**. Edge enhancement of letter "N". Left panel: original $8 \times 8$ binary image of letter "N". Middle and right panels: the simulation and experimental results.

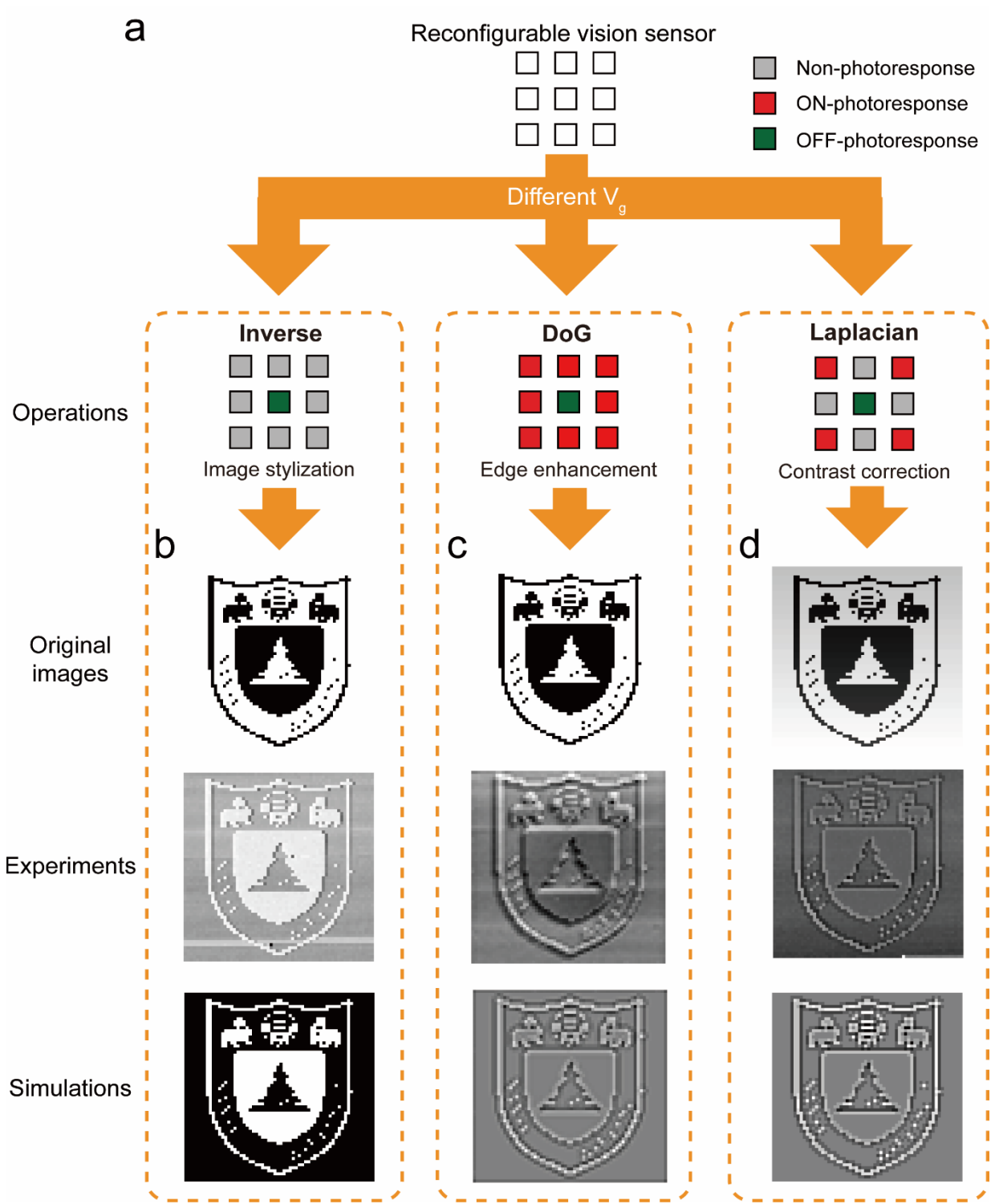

**Fig. 3 Reconfigurable retinomorphic vision sensor. a**. Demonstration of image processing with three different operations (*i.e.* image stylization, edge enhancement and contrast correction). These operations are realized by controlling the photoresponse of each pixel in the sensor by varying $V_g$ independently. **b**. Image stylization. **c**. Edge enhancement. **d**. Contrast correction. Original images correspond to the images to be processed by different operations. Experimental results by distinct convolution operations are compared with simulations.

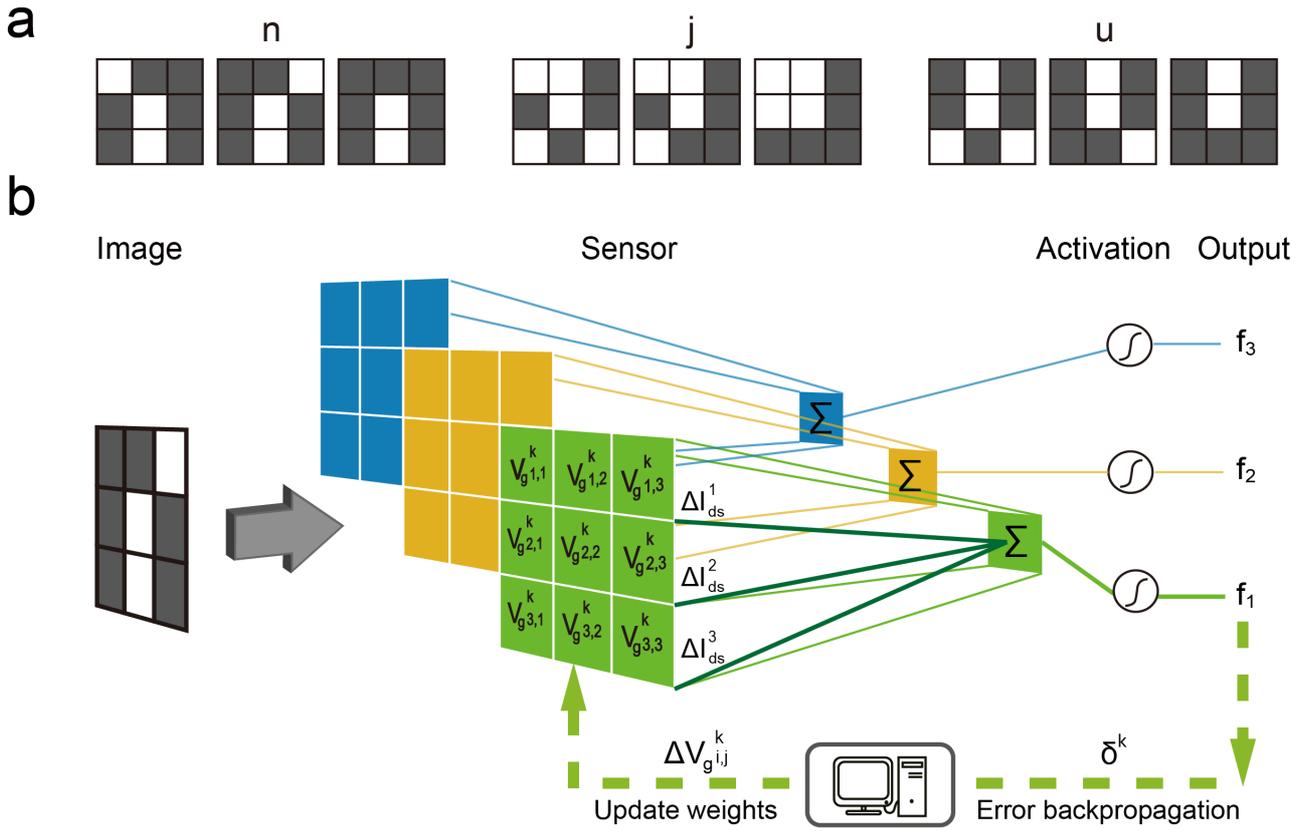
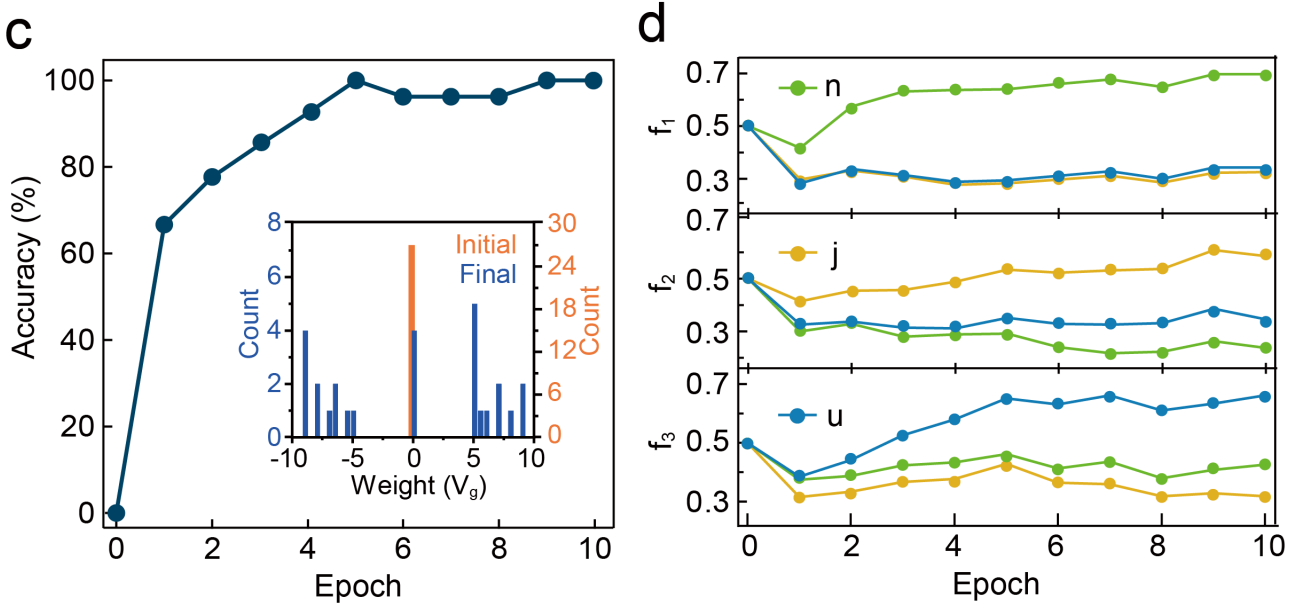

**Fig. 4 Implementation of convolutional neural network with the retinomorphic vision sensor. a.** Three different patterns of each specific letter ('n', 'j' and 'u'). **b.** Training process of the vision sensor at each epoch. The different colors maps correspond to different convolution kernels. $k$ is the number of training. $i$ and $j$ denote the location of each pixel in the sensor. **c.** Accuracy of recognition over the epoch, the inset shows the weight distribution of vision sensor, corresponding to initial (yellow) and final training (blue). **d.** Measured average output signals for each epoch for a specific input letter. The curves with largest values ($f_1$, $f_2$ and $f_3$, respectively) represent the recognition results of the target letters.